\pdfoutput=1

\documentclass{aa}  
\usepackage{answers}
\usepackage{natbib}
\usepackage{amsmath}
\usepackage{amssymb}
\usepackage{threeparttable}
\usepackage{gensymb}
\usepackage{url}

\newcommand{\FIG}[1]{}
\usepackage{graphicx}
\usepackage{txfonts}
\usepackage{natbib,twoopt}
\usepackage[breaklinks=true]{hyperref} 
\bibpunct{(}{)}{;}{a}{}{,} 
\makeatletter
\newcommandtwoopt{\citeads}[3][][]{\href{http://adsabs.harvard.edu/abs/#3}%
{\def\hyper@linkstart##1##2{}%
\let\hyper@linkend\@empty\citealp[#1][#2]{#3}}}
\newcommandtwoopt{\citepads}[3][][]{\href{http://adsabs.harvard.edu/abs/#3}%
{\def\hyper@linkstart##1##2{}%
\let\hyper@linkend\@empty\citep[#1][#2]{#3}}}
\newcommandtwoopt{\citetads}[3][][]{\href{http://adsabs.harvard.edu/abs/#3}%
{\def\hyper@linkstart##1##2{}%
\let\hyper@linkend\@empty\citet[#1][#2]{#3}}}
\newcommandtwoopt{\citeyearads}[3][][]%
{\href{http://adsabs.harvard.edu/abs/#3}
{\def\hyper@linkstart##1##2{}%
\let\hyper@linkend\@empty\citeyear[#1][#2]{#3}}}
\makeatother

\begin{document}

   \title{Modelling ripples in Orion with coupled dust dynamics and radiative transfer}

   \subtitle{}

   \author{T. Hendrix\inst{1}
          \and
          R. Keppens\inst{1}
          \and
          P. Camps\inst{2}
          }
          
   \institute{Centre for mathematical Plasma Astrophysics, Department of Mathematics, KU Leuven,\\ Celestijnenlaan 200B, 3001 Leuven, Belgium\\ \email{tom.hendrix@wis.kuleuven.be}  \and Sterrenkundig Observatorium, Universiteit Gent, Krijgslaan 281, B-9000 Gent, Belgi\"e}

   \date{}

 
  \abstract
   {}
   {In light of the recent detection of direct evidence for the formation of Kelvin-Helmholtz instabilities in the Orion nebula, we expand upon previous modelling efforts by numerically simulating the shear-flow driven gas and dust dynamics in locations where the H$_{II}$ region and the molecular cloud interact. We aim to directly confront the simulation results with the infrared observations.}
   {To numerically model the onset and full nonlinear development of the Kelvin-Helmholtz instability we take the setup proposed to interpret the observations, and adjust it to a full 3D hydrodynamical simulation that includes the dynamics of gas as well as dust. A dust grain distribution with sizes between 5-250 nm is used, exploiting the gas+dust module of the MPI-AMRVAC code, in which the dust species are represented by several pressureless dust fluids. The evolution of the model is followed well into the nonlinear phase. The output of these simulations is then used as input for the SKIRT dust radiative transfer code to obtain infrared images at several stages of the evolution, which can be compared to the observations.}
   {We confirm that a 3D Kelvin-Helmholtz instability is able to develop in the proposed setup, and that the formation of the instability is not inhibited by the addition of dust. Kelvin-Helmholtz billows form at the end of the linear phase, and synthetic observations of the billows show striking similarities to the infrared observations. It is pointed out that the high density dust regions preferentially collect on the flanks of the billows. To get agreement with the observed Kelvin-Helmholtz ripples, the assumed geometry between the background radiation, the billows and the observer is seen to be of critical importance.}
   {}


   \maketitle
%

\section{Introduction}

\begin{figure}
  \centering
  \centerline{\includegraphics[width=\columnwidth]{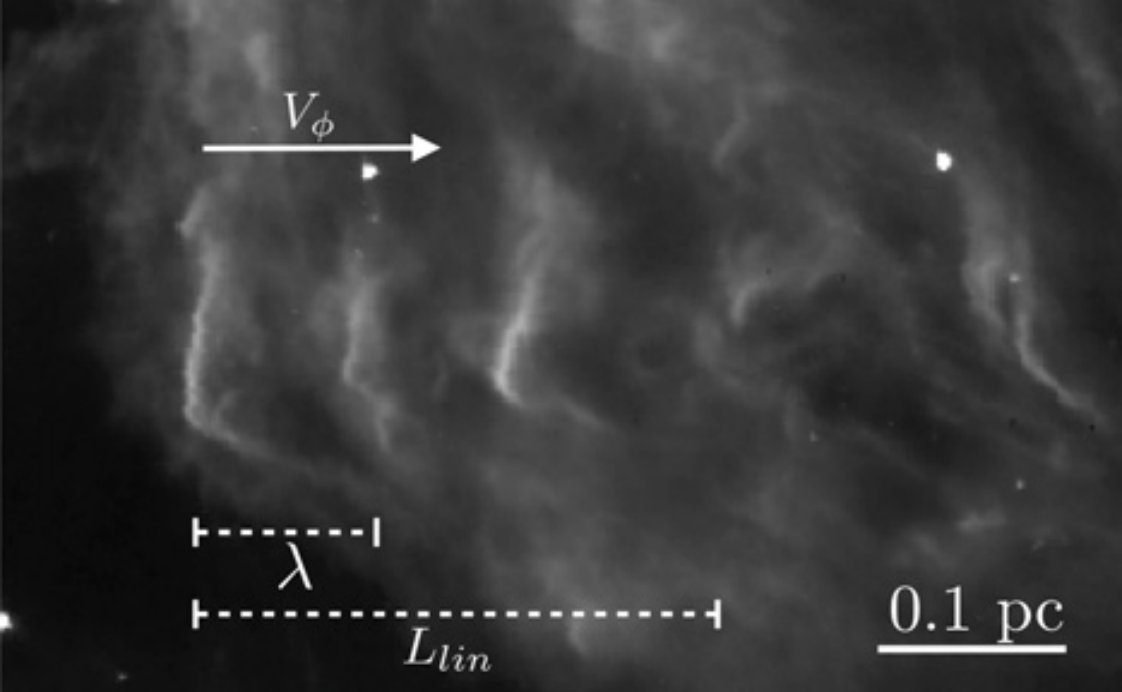}}
  \caption{Observation of the ripples in Orion at 8 $\mu$m, taken with the Spitzer Infrared Array Camera. The spatial wavelength $\lambda$, the orientation of the phase velocity $V_{\phi}$, and the linear regime length $L_{lin}$ are identified in the image. Credit: figure (1) from \citet{2012ApJ...761L...4B}, reproduced by permission of the AAS.}
\label{fig:berne}
\end{figure}

Sometimes a little push is all that is needed to make a seemingly stable fluid evolve into a turbulent state. Typically this transition is caused by a fluid instability, and many of these mechanisms have been studied extensively in the past decades (see e.g. 
\citet{1961hhs..book.....C}). The Kelvin-Helmholtz instability (KHI) is a notable example of this as it plays an important role in a wide range of different fluid applications such as for example oceanic circulation \citep{Haren}, winds on planet surfaces \citep{1997QJRMS.123.1433C}, the flanks of expanding coronal mass ejections \citep{2011ApJ...729L...8F}, magnetic reconnection in the solar corona \citep{2003SoPh..214..107L}, interaction between comet tails and the solar wind \citep{1980SSRv...25....3E}, mixing of solar wind material into Earth's magnetosphere \citep{2004Natur.430..755H}, astrophysical jets \citep{2006A&A...447....9B} and many others. While the KHI is a hydrodynamical instability, magnetic fields can alter its dynamics and cause stabilisation or further destabilise the setup. As the previous range of examples demonstrates, many of the relevant astrophysical fluids in the KHI is of importance display magnetic effects. In molecular clouds, the KHI has been linked to the formation of filamentary structures \citep{2014A&A...562A.114H}, as well as to turbulence formation. While the source of turbulence, observed in molecular clouds through the detection of non-thermal line-widths around $1\times10^5$ - $2\times10^5$ cm s$^{-1}$, is still debated, it has been linked at least partially to the KHI allowing to transfer energy to smaller scale structures \citep{2004ARA&A..42..211E,2011EAS....52..281B,2012ApJ...761L...4B}. While the occurrence of the KHI in space is clearly established, direct evidence of ongoing instabilities are harder to obtain. At a distance of 412 pc \citep{2009ApJ...700..137R}, the Orion nebula is the closest H$_{II}$ region. Its association with young massive stars and its apparent brightness make it an intensively investigated region over a large range of frequencies \citep{2001ARA&A..39...99O}. As such, it is an ideal laboratory for investigation of smaller scale structure development. Recently \citet{2010Natur.466..947B} discussed mid-infrared observations of ripple-like structures on the edge of the Orion nebula's H$_{II}$ region and the surrounding giant molecular clouds. The wave-like nature of this observation (see figure \ref{fig:berne}), points to a mechanism with fixed periodicity in time or space. This periodic structure, in combination with the detection of a strong velocity gradient resulting in velocity differences up to $7\times10^5$ - $9\times10^5$ cm s$^{-1}$ leads \citet{2010Natur.466..947B} to propose that these ripples are manifestations of the KHI. \\
Because of the high research interest in the Orion nebula and the surroundings regions, the physical conditions in the neighbourhood of the observed ripples are fairly well documented, providing an ideal case to numerically model the observed system. In \citet{2012ApJ...761L...4B} an effort was undertaken to numerically study the linear growth phase of a KHI with physical values deduced from observations. It was found that the used setup was indeed Kelvin-Helmholtz unstable for setups with magnetic field orientations close to perpendicular to the flow, and parallel to the separation layer between the H$_{II}$ and cloud region. \\
In this work, our goal is to expand the numerical modelling of the ripples in Orion in a way in which the observations can be directly compared to the modelling itself. To do so, several ingredients are needed. First, the proposed setup (see sections \ref{physSetup} and \ref{magp}) is simulated using a 3D numerical hydrodynamical simulation from the start of the instability, through the linear phase and into the nonlinear phase. To perform these simulations we use the {\tt MPI-AMRVAC} code \citep{2012JCoPh.231..718K,2014ApJS..214....4P}, with numerical properties as described in section \ref{NumMeth}. In the mid-infrared observation a significant part of the radiation is due to dust emission. Therefore we use the gas+dust module of the {\tt MPI-AMRVAC} code to model the dynamics of dust particles, which are drag-coupled to the gas. We use a range of dust sizes and model it self-consistently with the gas dynamics. Finally, to connect the dynamical simulations to the observations we use the {\tt SKIRT} dust radiative transfer code \citep{2011ApJS..196...22B,Camps201520} to emulate the radiation by the dust particles and the effect of the actual geometry of the observed system, as explained in section \ref{RadTrans}. The properties of the outcome of these simulations are described in section \ref{results} and the conclusions are discussed in section \ref{conclusions}.

\section{Model}

\subsection{Physical setup}
\label{physSetup}
The setup used here is similar to that of the 2D setup of \citet{2012ApJ...761L...4B}, but here adjusted to a full 3D configuration. The domain of the simulation is a cube with $L=0.33$ pc sides, and is initially divided in three regions along the $y$-axis: the upper part corresponds to the hot, low density H$_{II}$ region (n$_{II} = 3.34 \times 10^{-23}$ g cm$^{-3}$, T$_{II}$ = 10$^4$ K), the lower part represents the cold, high density molecular cloud (n$_c = 1.67 \times 10^{-20}$ g cm$^{-3}$, T$_{c} = 20$ K) and both are separated by a thin middle layer with thickness $D=0.01$ pc. This boundary layer is thus oriented perpendicular to the $y$-axis. Note that the choice of density and temperature result in thermal pressure equilibrium between the upper and lower region as
\begin{equation}
\qquad p = \rho \frac{k_b T}{m_H \mu},
\end{equation}
with $p$ the pressure, $k_b$ the Boltzmann constant, $m_H$ the mass of hydrogen and $\mu$ the average molecular weight, set to $\mu = 1$ here. The energy density of the gas , $e$, can be calculated using the equation of state, and gives
\begin{equation}
	\qquad e = \frac{p}{\gamma - 1} + \frac{\rho v^2}{2},
\end{equation}
with $\gamma = 5/3$ the adiabatic constant and $v$ the velocity of the flow.\\

To initialise the dust content in the simulation domain, we assume that the dust-to-gas mass density ratio has the canonical value of 0.01 \citep{1954ApJ...120....1S} in the molecular cloud region, and no dust is present in the hot H$_{II}$ region. We assume that the size distribution of dust particles, $n$, can be approximated as $n(a) \propto a^{-3.5}$ with the size of the particles, $a$, between 5 nm and 250 nm as was determined from excitation in the interstellar medium (ISM) by \citet{1994ApJ...422..164K}. We use four dust fluids to represent this power law size distribution with each fluid representing a part of the size distribution, chosen in a way in which the total dust mass in each dust fluid is the same (see \citet{2014A&A...562A.114H}). In this way, the resulting representative size of dust grain in the four dust fluids are 7.9 nm, 44.2 nm, 105 nm, and 189 nm, respectively. The grain density of all dust fluids is set to that of silicate grains, i.e. 3.3 g cm$^{-3}$ \citep{1984ApJ...285...89D}.\\
The H$_{II}$ region has an initially uniform velocity of magnitude $v_0 = 10^6$ cm s$^{-1}$ in the direction parallel to our $x$-axis. \citet{2012ApJ...761L...4B} propose that this high velocity is due to \textit{champagne flow}, the resulting high velocity flow when the expanding H$_{II}$ breaks trough the molecular cloud. This velocity is similar to the shear velocity derived from observation in \citet{2010Natur.466..947B}. In the molecular cloud region the velocity is initially set to zero. In contrast to \citet{2012ApJ...761L...4B}, where a hyperbolic tangent profile is used for both velocity and density, we use a linear profile in the middle layer that continuously links up with the constant velocities and densities on both sides of the layer. This is done in analogy with our previous work \citep{2014A&A...562A.114H}, as it allows to better quantify the linear stability properties.\\
A perturbation is added by introducing an initial velocity component perpendicular to the boundary layer:
\begin{align}
\qquad v_{y,0}(x,y,z) =&  10^{-3} v_0 \exp \left( -\frac{(y-M_y )^2}{2 \sigma_y^2} -\frac{(z-M_z)^2}{2 \sigma_z^2} \right) \sin{(k_x x)} \nonumber \\
\qquad + &10^{-4} v_0 \, \textrm{rect} (\frac{y}{5D}) (1 - 2\textrm{rand} ()) \label{perturb},
\end{align}
with $\sigma_y = 5D$, $\sigma_z = L / 5$ and $M_y$ and $M_z$ being the $y$- and $z$- coordinates of the middle point of the separation layer. The first part on the right side of equation (\ref{perturb}) adds a sine perturbation with wavelength $\lambda = k_x / 2\pi$. We adopt $\lambda = 0.11$ pc in accord with the observations in \citet{2010Natur.466..947B}. The second part 
on the right side of equation (\ref{perturb}) adds random velocities\footnote{The random function $\textrm{rand}$ generates a random floating point value between 0 and 1, while the $\textrm{rect}$ function (also called "rectangular function") is one between $-0.5$ and $0.5$ and zero elsewhere. } 
 between $-10^{-4} v_0$ and $10^{-4} v_0$ in a layer of thickness $5D$ around the middle of the separation layer. The velocity in the $z$-direction is seeded with a similar random term:
\begin{equation}
\qquad v_{z,0}(x,y,z) =  10^{-4} v_0 \, \textrm{rect} (\frac{y}{5D}) (1 - 2\textrm{rand} ()).
\end{equation}
The purpose of the exponential part in equation (\ref{perturb}) in the $y$-direction is to preferentially locate the perturbation around the middle layer. The exponential part in the $z$-direction centres the perturbation around the middle of the $z$-axis to confine the instability development region. These random perturbations in the velocity break the symmetry of the setup, and allow in essence all unstable modes to develop spontaneously, although the fixed $\lambda$ wavelength in the $x$-direction gets preference.

\subsection{Magnetic pressure}
\label{magp}

\citet{2012ApJ...761L...4B} take into account a magnetic contribution in their 2D setup as well, assuming a uniform magnetic field with a strength of $B = 200$ $\mu$G in the entire domain based on observations of surrounding regions \citep{2004ApJ...609..247A,2005ASPC..343..183B}. Using the values of the physical setup (section \ref{physSetup}) this results in a ratio between thermal and magnetic pressure $\beta_{pl} = p_t / p_M = 0.0173$, with $\beta_{pl} $ the plasma beta value, meaning that the magnetic pressure is dominant over the thermal pressure contribution. The dominance of magnetic over thermal pressure is confirmed by observations in the orion molecular cloud \citep{2014ApJ...795...13B}, both for large and small scale structures. \citet{2012ApJ...761L...4B} note that the setup is most unstable when the magnetic field is perpendicular to the flow and parallel to the contact layer. In this configuration, a uniform magnetic field only contributes as an additional magnetic pressure
\begin{equation}
\qquad p_M = \frac{B^2}{8\pi}.
\end{equation}
This means that one can actually substitute the full MHD treatment by a HD treatment with an additional pressure term, in which the total pressure is raised while keeping the density fixed (thus artificially increasing the temperature). When calculating the thermal energy of the gas to quantify the coupling to the dust (see \citep{2014ApJS..214....4P}), this artificial term is subtracted to obtain the relevant temperature. To demonstrate that this approximation is valid, we compare evolution of an MHD setup with that of a HD + $p_M$ simulation in section \ref{2Dcomp}.

\subsection{Numerical method}
\label{NumMeth}
We use the {\tt MPI-AMRVAC} code \citep{2012JCoPh.231..718K,2014ApJS..214....4P} for all the hydrodynamical (HD) and magnetohydrodynamical (MHD) simulations. The dust module of {\tt MPI-AMRVAC}, discussed in detail in \citet{2014A&A...562A.114H}, allows to add dust to a HD simulation by adding multiple dust fluids. These fluids follow the Euler equations with vanishing pressure \citep{rjl:dust} and couple to the gas fluid through a drag force term. Each dust fluid has its own physical properties such as grain size and grain material density. Typically we use multiple dust fluids with the same grain material density and different grain sizes to model the size distribution in the ISM. \\
For the 3D simulations we use four levels of adaptive mesh refinement (AMR), resulting in an effective resolution of $448\times 1792\times448$ cells. The triggering of extra refinement levels is based on a combination of the gradients in the gas fluid and those in the dust fluid representing the largest grains. Because the actual physical domain is cube shaped, this resolution results in a four time higher resolution perpendicular to the flow (see section \ref{physSetup}). This is necessary to resolve all small-scale variations that develop during the linear (and also the nonlinear) phase of the instability. The solution of the coupled gas+dust fluid equations is advanced using a total variation diminishing Lax-Friedrich (TVDLF) scheme with a two-step predictor-corrector time discretisation and a monotonised central (MC) type limiter \citep{1977JCoPh..23..263V}. To ensure stable time-stepping the timestep is limited by using a CFL number of 0.6 for gas and dust, as well a separate dust acceleration criterion based on the stopping time of dust grains \citep{2012MNRAS.420.2345L}.

\subsection{Radiative transfer}
\label{RadTrans}
To be able to directly compare the output from the 3D hydrodynamical simulations with observations, post-processing of the data is performed with the Monte Carlo radiative transfer code SKIRT \citep{2011ApJS..196...22B,Camps201520}. SKIRT simulates continuum radiation transfer in dusty astrophysical systems by launching a set of photon packages in a given wavelength range through the dust distribution obtained from our dynamical simulations. These packages are followed for several cycles of multiple anisotropic scattering, absorption and (re-)emission by interstellar dust, including non-local thermal equilibrium dust emission by transiently heated small grains. Emission from stochastically heated grains is used in all the results in this work and typically around 4 dust emission cycles are needed to come to equilibrium.\\
To launch the packages into the domain, we use a (stellar) point-source at a given distance outside of the simulated domain as our source of initial photons. Photon packages in a wavelength range between 0.01 $\mu$m and 1000 $\mu$m are incorporated. In SKIRT we use exactly the same distribution of dust species as the one obtained from MPI-AMRVAC, meaning that the mass density distribution of the four dust fluids is used for each representative part of the grain size distribution and that, just like in the HD simulations, we adopt silicate properties for the grains in the radiative transfer.

\section{Results}
\label{results}

\subsection{2D analysis}
\label{2Dcomp}

\begin{figure}
  \centering
  \centerline{\includegraphics[width=\columnwidth]{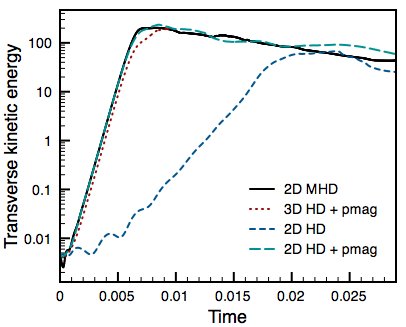}}
  \caption{Growth of the kinetic energy perpendicular to the bulk flow. The MHD and HD simulation that take into account the magnetic pressure are similar, while the HD simulation without magnetic pressure behaves differently. The 3D setup is also shown up to $t=0.01$ and has a growth rate similar to that of the 2D setup.}
\label{fig:linGrowth}
\end{figure}

\begin{figure}
  \centering
  \centerline{\includegraphics[width=\columnwidth]{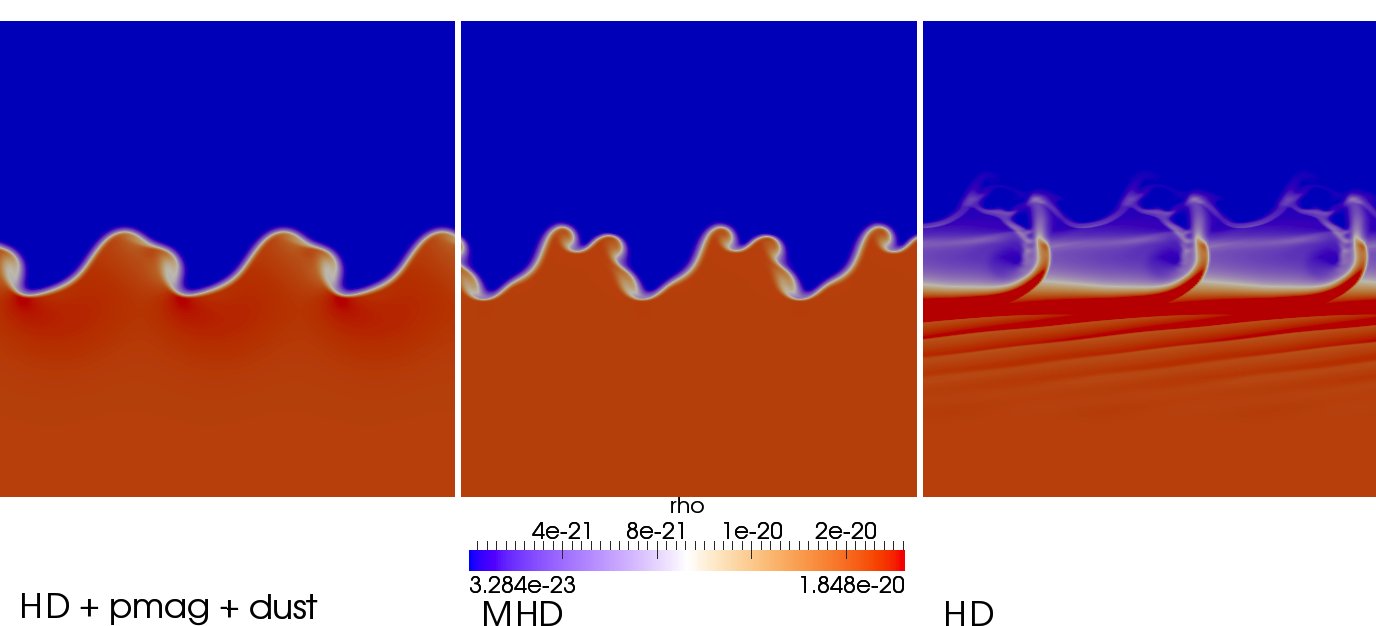}}
  \caption{Gas density plots of the KHI in 2D and 2.5D after the end of the linear phase. The density units are in g cm$^{-3}$. In all figures the entire domain (0.33 pc $\times$ 0.33 pc) is shown. \textbf{Left:} A 2D simulation of the KHI in HD with dust and an artificial magnetic pressure term $p_M$ added to the total pressure at $t=0.007$  (6.84$\times10^4$ years). \textbf{Centre:} The same setup, but in 2.5D MHD with a magnetic field perpendicular to the plane, also at $t=0.007$. \textbf{Right:} A 2D HD simulation without the effect of a magnetic field added into the total gas pressure, at $t=0.02$ (1.95$\times10^5$ years). Note this figure is taken at a different time as the linear phase end later in this case.}
\label{fig:magNomag}
\end{figure}

To prove that an MHD setup with the magnetic field component perpendicular to the flow direction and parallel to the boundary layer can be reasonably approximated by a similar setup in HD but with added pressure, we simulate the setup discussed in sections \ref{physSetup} and \ref{magp} first in 2D, but in three variations: a HD simulation without a magnetic contribution, an MHD setup with magnetic field, and an HD simulation with the magnetic field contribution added to the pressure. The MHD setup is actually simulated in 2.5D, as it includes the information of the velocity and magnetic field perpendicular to the simulated plane. The simulated plane in 2D corresponds to a slice in the 3D simulation perpendicular to the $x-y$ plane and through the centre of the simulated domain. In figure \ref{fig:linGrowth} the buildup of kinetic energy perpendicular to the flow direction is shown for all three 2D setups, and for the 3D run discussed further on. Clearly, for the MHD setup and the HD plus magnetic pressure setup the growth rate in the linear regime (up to $t=0.006$ in code units, or $\sim$ 5.87$\times10^4$ years) is the same. The growth rate is significantly slower when the magnetic pressure is ignored. Also, figure \ref{fig:magNomag} shows that the formed structures are of similar size and shape in the two simulations where the magnetic pressure is taken into account. Small differences include the formation of small-scale structures on top of the larger structure. These small-scale perturbations are also present in the HD setup, but develop faster in the MHD simulation. The reason that they are less apparent in the HD simulation is because in the MHD case they seemingly grow faster due to small inhomogeneities (a decrease by $\approx 2\%$) in the magnetic field, leading to numerical differences that accumulate over time. When the magnetic pressure is not taken into account, it can be seen in figure \ref{fig:magNomag} that the morphology is very different. Because the total pressure is lower, the Mach number for the flow at the boundary is higher, causing shocks to propagate. These shocks also cause the striped structure in the high density region. We will now further discuss a full 3D gas plus dust setup that has the pressure adjusted to account for the magnetic pressure effects.

\subsection{3D model}

In figure~\ref{fig:linGrowth} it can be seen that the growth rate of the 3D simulation is comparable to that of the 2D simulations in which the effect of the magnetic field is taken into account. Due to the added computational cost in 3D, this simulation is only followed until $t=0.01$ in code units, or up to about 9.78$\times10^4$ year.

\subsubsection{Dust distribution}
\label{dustDistri}

\begin{figure}
  \centering
  \centerline{\includegraphics[width=0.9\columnwidth]{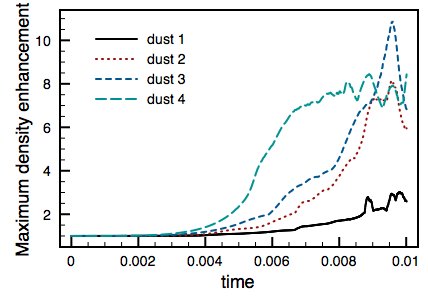}}
  \caption{Time evolution of the maximal density enhancements in the 3D simulation for all four dust fluids, with \textit{dust 1} representing the smallest grains (7.9 nm) and \textit{dust 4} the largest grains (189 nm).}
\label{fig:maxDens}
\end{figure}

In previous work \citep{2014A&A...562A.114H} we found that in a 3D setup with the same density on both sides of the separation layer, the KHI can cause the dust density to increase by almost two orders of magnitude. These strong increases in dust density occur in filament-like locations between the vortices when dust is swirled out of the vortices and compressed into these regions. This process if strengthened further by additional 3D instabilities. Also, it was found that the process of dust density enhancement is stronger for larger dust particle sizes. Figure \ref{fig:maxDens} shows that in the setup used here the growth in local dust density is less strong. During the end of the linear phase, i.e. up to time $t=0.006$ in figure~\ref{fig:maxDens}, the maximal density increases gradually, and the rate of increase is proportional to the grain size. In the further nonlinear stage the densities still increase, however the relation between instantaneous local maximal density and grain size gets modified. Similarly to what was seen in \citet{2014A&A...562A.114H}, the density enhancements are significantly stronger in 3D than in 2D, where the maximum increase is less than $15\%$ for all dust species in the 2D case with magnetic pressure added. Clearly, 3D effects are paramount when studying dust growth.\\

\begin{figure}
  \centering
  \centerline{\includegraphics[width=0.8\columnwidth]{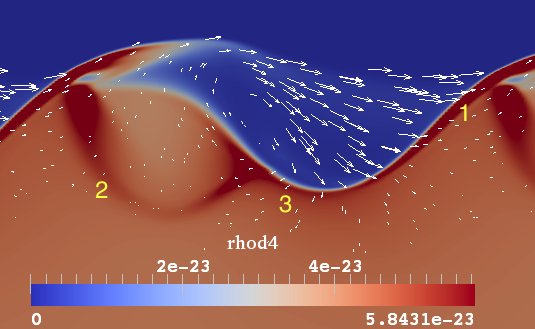}}
  \caption{Density of the largest dust species ($a=189$ nm) in a slice from the 3D simulation ($z=0.165$pc) at $t=0.0065$ (6.36$\times10^4$ years). Only a part of the simulated region with an extend of 0.138 pc in the $x$-direction is shown. Three distinct regions of dust density enhancement are indicated with labels 1, 2 and 3 discussed in the text. The velocity field of the largest dust species in the $x-y$ plane is indicted with the use of vectors, the largest velocity are around $6 \times 10^5$ cm s$^{-1}$.}
\label{fig:regions}
\end{figure}

\begin{figure}
  \centering
  \centerline{\includegraphics[width=\columnwidth]{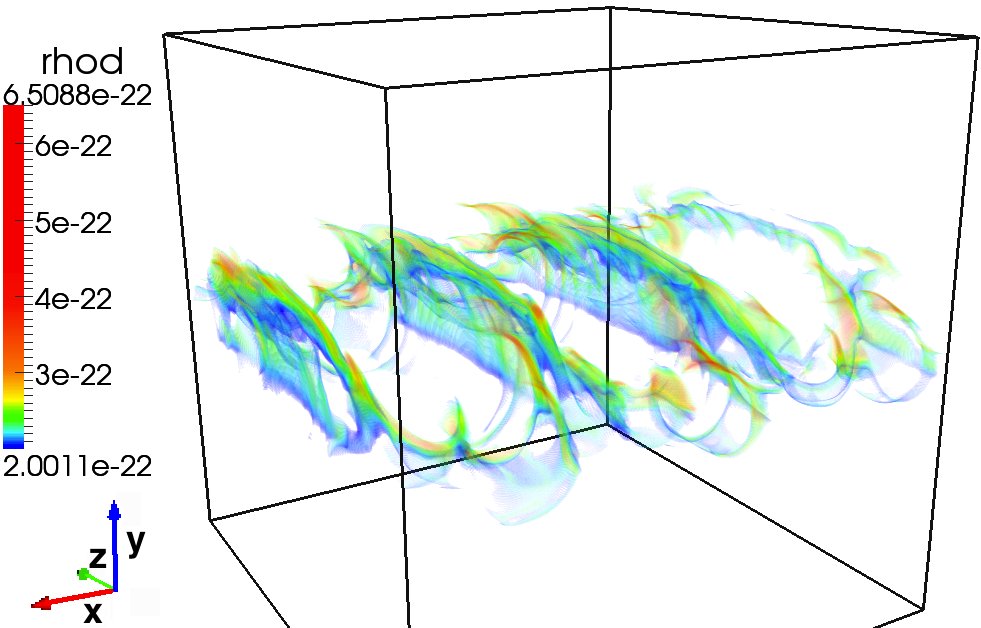}}
  \caption{Volume plot of the total dust density at $t=0.01$ (9.78$\times10^4$ years). Only densities higher than the initial maximum density ($\rho_d = 1.67\times 10^{-22}$ g cm$^{-3}$) are visualised.}
\label{fig:rhodTot}
\end{figure}

\noindent The dust density enhancements are strongest in three distinct regions, which are indicated in figure \ref{fig:regions}. Chronologically dust first accumulates in the convex outer region of the KH wave (the region labeled with 1 in figure \ref{fig:regions}). This is due to the acceleration of dust by gas in the concave region when the gas swirls around the low pressure region created by the KHI. Next, the arc-like structure below the surface of the wave, i.e. region number 2 in figure \ref{fig:regions}, is formed. This region forms when the KHI accelerates the bulk of the gas upward into the low density region, and the dust is dragged with it. The location of the region is caused by a gradient in the drag strength, as the velocity difference between gas and dust is stronger under the region than above, causing the underlying dust to overtake the dust above it. The third dust gathering region is along the boundary between high and low density regions in between two successive waves or KHI rolls. A dust pile-up is seen here in the nonlinear stage when the velocity of the gas around the low pressure vortex is highest. In animated views one can see how the end point of the flow that passes over the crest of the waves moves from location~1 to a spread out region all along the density boundary, i.e. up to location 3 as indicated. \\
While dust density increases up to a factor 10 are observed in these three regions for the four dust species, the actual location of these dust-gathering regions does not necessarily fully coincide for all dust species, similar to the findings in~\cite{2014A&A...562A.114H} where a clear size-separation was evident. Also, the actual importance of the three regions is distinct for different grain sizes. Therefore, the increase of the total dust density will be less strong and distributed over a larger region. Furthermore, the strongest increases can be found in small local clumps, as can be seen in figure \ref{fig:rhodTot}, visualising the total dust density concentrations. Quantitatively speaking, while 14.76$\%$ of the total volume experiences a total dust density enhancement of more than 5$\%$, in only 0.03$\%$ of the total volume the total dust density more than doubles (regions indicated in orange and red in figure \ref{fig:rhodTot}). This is in contrast with the 3D simulations in \citet{2014A&A...562A.114H}, where the high density dust is found in long filamentary structures and more than 4.5 $\%$ of the volume exhibits a doubling of the total dust density. The main differences reside in the adopted initial density contrast, as well as the fact that here only the molecular cloud region initially had dust. 

\subsection{Modelling observations}

\begin{figure}
  \centering
  \centerline{\includegraphics[width=\columnwidth]{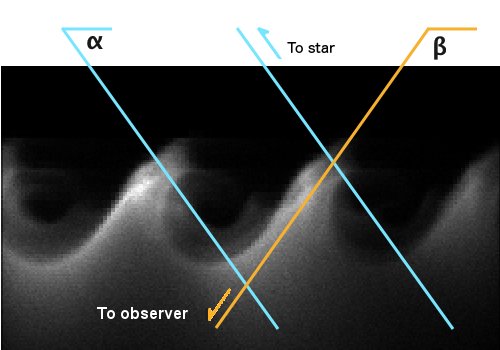}}
  \caption{Geometry of the stellar object (photon source) and observer location with respect to the structures in Orion, designated by independent angles $\alpha$ and $\beta$, respectively. In this image, the location of the source and observer are shown with respect to the KH features at t=0.084 (8.21$\times10^4$ years). The black-white image is actually a SKIRT image at 54 $\mu$m, where we see the radiation which is coming from dense and heated dust in the billow structures formed by the KHI. In this image, the observer is located perpendicular to the $x-y$ plane.}
\label{fig:geo}
\end{figure}

In the previous section we have outlined how the model setup from section \ref{physSetup} evolves into a nonlinear 3D KHI. Next, we investigate how the simulated structures would look in synthetic observations. As described in section \ref{RadTrans}, the dust distribution of our 3D simulations is used as input for the SKIRT radiative transfer code. To see to which degree our simulations correspond to the actual observed structures (figure \ref{fig:berne}), in addition to the hydrodynamical setup one has to take into account the orientation in relation to the observer, as well as the location of the light source(s). \citet{2010Natur.466..947B} indicated that the star $\theta^1$ Orionis C, a massive type O7V star \citep{2002MNRAS.333...55D,2006A&A...451..195W} located in the H$_{II}$ Trapezium region at a distance of $\sim$ 3.4 pc from the cloud, illuminates the ripples from behind with respect to the observer. In SKIRT the radiation of this star is simulated by adding a point source of photons at $d=3.4$ pc and inclination $\alpha$ with respect to the initial separation layer in the HD simulation, as illustrated in figure \ref{fig:geo}. For the radiation of the star we use a model spectrum from \citet{2005A&A...436.1049M} with corresponds to a star with physical properties comparable to those of $\theta^1$ Orionis C\footnote{Model T46p1\_logg4p05.sed from \url{http://www.mpe.mpg.de/~martins/SED.html}}. The location of the observer with respect to the simulated domain must also be specified in SKIRT. As shown in figure \ref{fig:geo}, the observer is placed at an angle $\beta$ with respect to the initial separation layer in the HD simulation. \\

\begin{figure}
  \centering
  \centerline{\includegraphics[width=1.03\columnwidth]{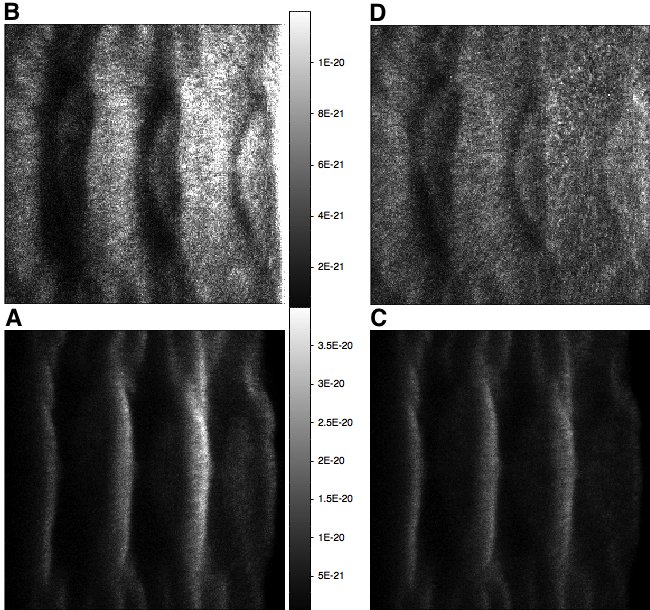}}
  \caption{SKIRT simulations of the same dataset with different geometries. From left to right and top to bottom: B, D, A, C. Horizontally the observers angle $\beta$ is the same ($\beta = 90^{\degree}$ on top, $\beta = 128^{\degree}$ below) and the same scaling is used. Note that the flux quantification is arbitrary here and no effort has been taken to compare these to real values. Vertically the irradiation angle is constant ($\alpha = 40^{\degree}$ left, $\alpha = 60^{\degree}$ right). All images are observed at 8.25 $\mu$m. }
\label{fig:BDAC}
\end{figure}

\begin{figure}
  \centering
  \centerline{\includegraphics[width=1.05\columnwidth]{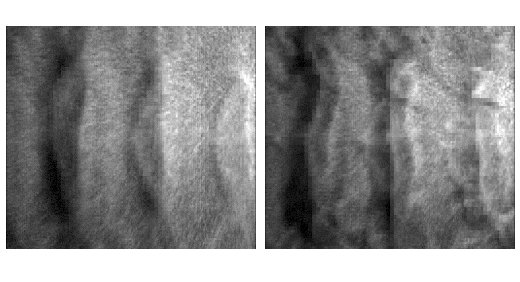}}
  \caption{Synthetic observation of the KHI at 8.25 $\mu$m, with fixed observational angle $\beta=90^{\degree}$ and $\alpha=128^{\degree}$ (cases E and G). Two different times are shown, left: $t=0.0084$, right: $t=0.01$ or 8.21$\times10^4$ and 9.78$\times10^4$ year, respectively). During this interval the development of small-scale perturbations in the nonlinear phase can be seen. A linear scale is used for the intensity of the images.}
\label{fig:perp}
\end{figure}

Because the actual inclination between the observer, the billows and the background radiation source are hard to gauge from the observation, several different values of $\alpha$ and $\beta$ were tried to investigate their role. Table \ref{table:1} gives an overview of several SKIRT geometries we will discuss here. An interesting setup to look at first is case D (figure \ref{fig:BDAC}, top right). With this arbitrary choice for the geometry ($\alpha=60^{\degree}$ and $\beta=90^{\degree}$) the result is rather different from the observations. While some periodicity is observable, no sharp elongated structures are seen. The diffuseness of the radiation in case D can be seen to be inherent to an observer angle of $90^{\degree}$. Figure \ref{fig:perp} demonstrates that when going from $t=0.0082$ in E to $t=0.01$ in G, while the onset of the nonlinear phase increases the development of small-scale features (as discussed in section \ref{dustDistri}), the emission in the nonlinear phase remains diffuse in both cases.\\

\begin{figure}
  \centering
  \centerline{\includegraphics[width=1.05\columnwidth]{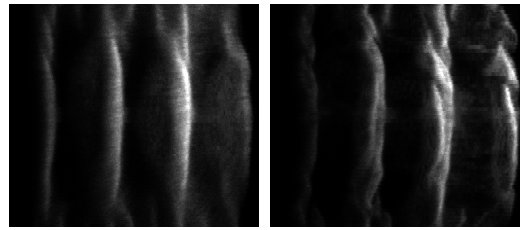}}
  \caption{Synthetic observation of the KHI at 8.25 $\mu$m, with observational angle $\beta=128^{\degree}$ and $\alpha=51^{\degree}$ (cases F and H). Two different times are shown, left: $t=0.0084$, right: $t=0.01$ or 8.21$\times10^4$ and 9.78$\times10^4$ year, respectively). In comparison to the images at $\beta=90^{\degree}$, the features of the KHI are more pronounced and clearly distinguishable from the background. A linear scale is used for the intensity of the images. }
\label{fig:opti}
\end{figure}

In figure \ref{fig:geo} we see that the emission at 54 $\mu$m is strongest where the dust is directly radiated by the source, but the colder dust inside the KH billows also radiates at this wavelength. At shorter wavelengths such as 8.25 $\mu$m, the direct light is the more important and only dust close to the edges of the billows radiates. To get features more reminiscent of the observations we can use this knowledge to consider two changes to the geometry of the source and the observer. On the one hand, the angle $\alpha$ can be chosen to maximise the photons from the source reaching the protruding billows and not the rest of the cloud, which increases the amount of observed photons in a more compact location. Nevertheless, the effect of changing $\alpha$ is small at 8.25 $\mu$m, as demonstrated by comparing cases A to C and B to D in figure \ref{fig:BDAC}. On the other hand the observers angle $\beta$ can be chosen to be along the billows, maximising the perceived compactness. The change in observer angle has a much stronger impact. Changing $\beta$ from $90^{\degree}$ in case B to $\beta=128^{\degree}$ in case A clearly decreases the thickness of the features, increases the flux in the elongated regions, and enhances the contrast between the bright en dark regions. The choice for ``optimal angles" is illustrated in figure \ref{fig:geo}. The values we find are $\alpha=51^{\degree}$ and $\beta=128^{\degree}$. These values are used in cases F and H (figure \ref{fig:opti}). Using this geometry, a fair approximation of the real observations can be made, at a comparable wavelength. The evolution from case F into H again displays the formation of the small scale structures in the nonlinear phase, on a scale which is comparable to the local bends in the infrared observations.

\begin{table}
\caption{Summary of the SKIRT radiative transfer models, with $\alpha$ the angle between the star and the cloud, and $\beta$ the angle between the cloud and the observer (see figure \ref{fig:geo}), and the time in code units.}              
\label{table:1}      
\centering 
\begin{tabular}{c c c c}
Case       & $\alpha$ & $\beta$  & time \\
\hline
A & 40 & 128 & 0.0082\\
B & 40 & 90 & 0.0082\\
C & 60 & 128 & 0.0082\\
D & 60 & 90 & 0.0082\\
E & 51 & 90 & 0.0082\\
F & 51 & 128 & 0.0082\\
G & 51 & 90 & 0.01\\
H & 51 & 128 & 0.01\\
\end{tabular}
\end{table}



\section{Conclusions}
\label{conclusions}

In the previous sections, we have modelled a region of the Orion molecular cloud in which elongated ripple features are observed. To do so, we have built upon previous numerical models, and expanded these to full 3D dusty hydrodynamics coupled to a radiation transfer code designed for simulating dusty astrophysical systems. The synthetic images allow a direct comparison with the observations. In the infrared observations, the ripples are thin, elongated features that have a clear periodicity and are sharp and bright compared to the background radiation. All these features can also be reproduced by our model. The hydrodynamical simulations confirm that the previously proposed setup is indeed KH unstable for the observed spatial wavelength. We find that the dynamical contribution of dust with a size distribution typical for the ISM does not inhibit the formation of the KHI, and the growth rate in 3D is similar to that of the 2D simulation. We see that the presence of a background star is able to light up the features of the KH billows. Also, the synthetic images demonstrate clearly that the geometry is of great importance in distinguishing the KH features from the background. Observers located in a direction perpendicular to the shearing layer would observe some periodicity, however with shallow features over a continuous background, while observers which look along the formed billows observe them very sharp and bright compared to the background. Nevertheless, even when considering the most optimal geometry, the ripples are still somewhat wider than the sharp ripples of the observations. Additional to geometrical effects, the sharp features may point to strong local density increases in the dust, however in contrast to our previous investigation of dusty KHI \citep{2014A&A...562A.114H} only small increases in dust density are seen here, and the highest increases are found in small and compact clumps and not elongated regions. The treatment of additional physics such as self gravity and magnetic fields may lead to these additional density increases as was shown for larger scale structures in \citet{2014ApJ...789...37V}. It is unclear if a significant effect would also be expected here, as in section \ref{2Dcomp} the magnetic field only causes minor deviations in the 2D setup. For simulations in 3D, the strong magnetic field (plasma $\beta_{pl}=0.0173$) may somewhat alter the outcome of the simulations in the nonlinear phase, when secondary 3D instabilities break the earlier quasi-2D behaviour. \citet{2000ApJ...545..475R} demonstrated that even weak magnetic fields can be of importance in the nonlinear regime. While a strong magnetic fields may suppress the growth of hydrodynamical perturbations perpendicular to the fields, \citet{2007JGRA..112.6223M} find that in cases with plasma beta as low as $\beta_{pl}=0.1$ secondary 3D instabilities also occur and cause small scale fragmentation along the initial magnetic field, however at a stage far in the nonlinear regime. The resulting influence of the 3D magnetic field on the dynamics of the dust grains, and thus also the observed structures, is further complicated by the unknown charge of the dust grains. While for example \citet{2012ApJ...747...54H} have calculated mean grain charging as function of grain sizes for different ISM phases, the charging of grains can be location dependant due to for example interaction with a radiation field, as is the case here. Fully taking into account the magnetic field would thus also require further assumptions to be made with regard to dust distribution as a function of the both the size and the charge. Furthermore, the strength of the magnetic field is one of the less constrained parameters in the model; while the value in the model ($B = 200$ $\mu$G) is representative for surrounding regions, no local measurements of orientation and strength exist to our knowledge. As the magnetic pressure is shown to be of importance in finding the correct value for the growth rate (section \ref{2Dcomp}), the outcome would be different if a different magnetic field was assumed. This would especially be the case for different relative orientations of this field and the flow shear. \\

Another important factor which may change the outcome of the simulations is the actual width of the shearing layer between the hot medium and the molecular cloud.
The width is an important parameter in the evaluation of the stability and growth of the KHI instability. The value used here ($D = 0.01$ pc) is in analogy with the value of \citet{2012ApJ...761L...4B} where it is argued that this value represents the width of the photodissociation region (PDR), where molecular gas is dissociated by the far ultraviolet photons of the background star $\theta^1$ Orionis C. Nevertheless, as discussed in the supplement of \citet{2010Natur.466..947B}, actually a broader ($\sim 0.1$ pc) photo-ablation region forms between the PDR and the hot medium. Due to its thickness this region may inhibit the formation of the KHI with wavelengths in the range of the observed periodicity in the ripples or shorter, as a boundary layer of thickness $D$ inhibits the growth of perturbations with $\lambda < 4.91 D$ \citep{1961hhs..book.....C,2014A&A...562A.114H}. Additionally it should be noted that the effect of heat conduction, which has not been included in this work, can be of importance in the formation of the shearing layer between the hot medium and the molecular cloud. Indeed, \citet{2007A&A...472..141V} demonstrate that heat conduction can reduce the steepness of the velocity gradient between the cloud and a streaming flow, stabilising the surface of the cloud against the development of the KHI. \\

While these remarks demonstrate that additional physics may be needed to understand the full range of interactions occurring in the Orion nebula, in this work we tried to model the observations of its KH ripples in full detail. We demonstrated that a full treatment of gas and dust dynamics, including a range of dust sizes, coupled with radiative transfer provides a promising approach to explaining the observations. Even though the physical values in the models are prone to intrinsic observational uncertainties or assumptions, we see that these values are reasonable in reproducing most of the features when the most optimal geometrical model is used.

\begin{acknowledgements}
We acknowledge financial support from project GOA/2015-014 (KU Leuven) and by the Interuniversity Attraction Poles
Programme initiated by the Belgian Science Policy Office (IAP P7/08
CHARM). Part of the simulations used the infrastructure of the VSC -
Flemish Supercomputer Center, funded by the Hercules Foundation and the
Flemish Government - Department EWI. 
\end{acknowledgements}

\bibliography{aa25498-14}

\end{document}